# Chemical and Lattice Stability of the Tin Sulfides

Jonathan M. Skelton,[1*] Lee A. Burton,[2] Fumiyasu Oba,[2] and Aron Walsh[1,3,4]

[1] *Department of Chemistry, University of Bath, Claverton Down, Bath BA2 7AY, UK*

[2] *Laboratory for Materials and Structures, Institute of Innovative Research, Tokyo Institute of Technology, 4259 Nagatsuta, Midori-ku, Yokohama 226-8503, Japan*

[3] *Department of Materials, Imperial College London, Exhibition Road, London SW7 2AZ, UK*

[4] *Global E³ Institute and Department of Materials Science and Engineering, Yonsei University, Seoul 120-749, Korea*

[*] *Corresponding author; e-mail: j.m.skelton@bath.ac.uk*

**Abstract**

The tin sulfides represent a materials platform for earth-abundant semiconductor technologies. We present a first-principles study of the five known and proposed phases of SnS together with $SnS_2$ and $Sn_2S_3$. Lattice-dynamics techniques are used to evaluate the dynamical stability and temperature-dependent thermodynamic free energy, and we also consider the effect of dispersion forces on the energetics. The recently identified π-cubic phase of SnS is found to be metastable with respect to the well-known orthorhombic *Pnma*/*Cmcm* equilibrium. The *Cmcm* phase is a low-lying saddle point between *Pnma* local minima on the potential-energy surface, and is observed as an average structure at high temperatures. Bulk rocksalt and zincblende phases are found to be dynamically unstable, and we show that whereas rocksalt SnS can potentially be stabilised under a reduction of the lattice constant, the hypothetical zincblende phase proposed in several earlier studies is extremely unlikely to form. We also investigate the stability of $Sn_2S_3$ with respect to SnS and $SnS_2$, and find that both dispersion forces and vibrational contributions to the free energy are required to explain its experimentally-observed resistance to decomposition.

**Introduction**

The tin sulfides are a technologically-important family of earth-abundant optoelectronic materials comprising tin monosulfide (SnS), tin disulfide ($SnS_2$) and tin sesquisulfide ($Sn_2S_3$). SnS has long been explored as a sustainable candidate photovoltaic (PV) absorber material.[1-3] $SnS_2$ is a two-dimensional semiconductor that can be exfoliated to give individual nano-sheets,[4] and $Sn_2S_3$ is an example of a comparatively rare mixed-valence binary compound, possesses an unusual one-dimensional bonding structure,[5] and is predicted to show ambipolar dopability.[6]

The $Sn_xS_y$ system has a rich phase space.[7] SnS is known to form several (meta)stable phases, which has led to a degree of uncertainty over which phase(s) are obtained under different growth conditions. There are currently five reported or proposed phases, *viz.* the ground-state orthorhombic *Pnma* phase,[8] a high-temperature *Cmcm* phase,[9] and three cubic phases - rocksalt,[10] zincblende[11-13] and the recently-reported $P2_13$ ("π-cubic") phase with a 64-atom primitive cell.[14-17] The sesquisulfide $Sn_2S_3$ has also been a source of confusion, as it is relatively easy to prepare, and clearly distinguishable from the other tin sulphides,[18] yet is frequently predicted to be unstable with respect to decomposition into SnS and $SnS_2$ in theoretical studies (e.g. as in the current Materials Project[19] entry, mp-1509[20]). Both issues are important for contemporary PV research, since not only are phase impurities highly likely to play a role in the underwhelming performance of current SnS-based devices,[18, 21] but tin sulfides may also form as impurities during the growth of more complex multicomponent semiconductors such as $Cu_2ZnSnS_4$ (CZTS).[22]

First-principles materials modelling, e.g. within the ubiquitous Kohn-Sham density-functional theory formalism,[23, 24] affords a versatile means of exploring the subtleties of the equilibria between competing phases. However, to enumerate such a complex phase space, two common approximations made in contemporary modelling studies may need to be revisited, namely: (1) the poor description of weak dispersion interactions afforded by most generalised-gradient approximation (GGA) density functionals, which may be an issue in describing the sulfide phases with prominent non-bonding interactions;[5, 25] and (2) the omission of vibrational contributions to the temperature-dependent free energy in thermodynamic models.[26, 27] Consideration of the lattice dynamics would also enable the assessment of the bulk dynamical stability of the different phases, providing important insight into which phases are likely or unlikely to form under typical growth conditions.

In this work, we address both issues through a consistent set of first-principles lattice-dynamics calculations on the five known and proposed phases of SnS together with $SnS_2$ and $Sn_2S_3$, performed using both the PBEsol GGA[28] and the dispersion-corrected[29] PBEsol + D3 functionals. We establish the dynamical and thermodynamic stability ordering of the competing SnS phases, and provide conclusive evidence that the various experimental reports of cubic SnS are likely to be the rocksalt or π-cubic phases rather than a zincblende polymorph. We also consider the thermodynamics of the decomposition of $Sn_2S_3$, and show that to predict its stability requires an accurate description of the weak interactions in its structure as well as accounting for the vibrational contributions to its free energy.

**Computational methods**

First-principles calculations were carried out using the pseudopotential plane-wave density-functional theory (DFT) formalism, as implemented in the Vienna *Ab initio* Simulation Package (VASP) code.[30]

We performed calculations using the PBEsol functional[28] and PBEsol with the DFT-D3 dispersion correction[29] applied (i.e. PBEsol + D3). PBEsol was selected due to its being shown in a number of studies to provide a good description of the structural and vibrational properties of solids at a moderate computational cost.[31-33]

Initial models built from published crystal structures[8, 9, 15, 34-37] were fully optimised to a tolerance of $10^{-2}$ eV Å$^{-1}$ on the forces, using a plane-wave cutoff of 550 eV and carefully-converged **k**-point sampling (Table 1). A tight energy-convergence criterion of $10^{-8}$ eV was applied during the electronic minimisation. Projector-augmented-wave (PAW) pseudopotentials[38, 39] were used to model the core electrons, with the Sn 5s, 5p and 4d and the S 3s and 3p electrons being included in the valence shell.

| System | Phonon Supercell (# Atoms) | k-points Optimisation | k-points Phonon |
|---|---|---|---|
| $\alpha$-Sn | - | 12×12×12 Γ-MP | - |
| $\beta$-Sn | - | 16×16×32 Γ-MP | - |
| S | - | 3×1×1 Γ-MP | - |
| SnS (*Pnma*) | 6×1×6 (384) | 8×4×8 MP | 2×4×2 MP |
| SnS (Cubic) | 2×2×2 (512) | 2×2×2 Γ-MP | 1×1×1 Γ-MP |
| SnS (*Cmcm*) | 6×1×6 (384) | 8×4×8 MP | 2×4×2 MP |
| SnS (Rocksalt) | 4×4×4 (128) | 12×12×12 Γ-MP | 3×3×3 Γ-MP |
| SnS (Zincblende) | 4×4×4 (128) | 8×8×8 Γ-MP | 2×2×2 Γ-MP |
| SnS$_2$ | 6×6×2 (144) | 8×8×6 Γ-MP | 2×2×3 Γ-MP |
| Sn$_2$S$_3$ | 2×4×2 (320) | 4×8×3 Γ-MP | 2×2×2 Γ-MP |

**Table 1** Phonon supercells and **k**-point sampling used for the geometry optimisations and supercell finite-displacement phonon calculations performed in this work (MP - Monkhorst-Pack mesh,[40] Γ-MP - Γ-centered MP mesh).

The PAW projection was performed in reciprocal space. Non-spherical contributions to the gradient corrections inside the PAW spheres were accounted for, and the precision of the charge-density grids was set automatically to avoid aliasing errors.

Lattice-dynamics calculations were performed on the optimised structures using the Phonopy package,[41, 42] which was used to set up and post process supercell finite-displacement phonon calculations.[43] VASP was used as the force calculator, and an additional charge-density grid containing 8× the number of points as the standard one was used to evaluate the augmentation charges. The supercell expansions used to determine the second-order force-constant matrices are listed in Table 1. For all models apart from *Cmcm* SnS, the force constants were calculated in expansions of the primitive unit cells. The force constants for *Cmcm* SnS were calculated in an expansion of the conventional cell, and a transformation matrix to the primitive cell was applied when evaluating its phonon dispersion curves.

Phonon density of states (DoS) curves were obtained by interpolating the phonon frequencies onto a uniform 48×48×48 Γ-centered **q**-point mesh and using the linear tetrahedron method for the Brillouin-zone integration. The phonon dispersions were obtained by interpolating the phonon frequencies along lines of **q**-points passing between the high-symmetry points in the Brillouin zones of the primitive unit cells. Thermodynamic functions were computed from a set of phonon frequencies evaluated on the same 48×48×48 **q**-point grids as were used to obtain the phonon DoS curves.

**Results & Discussion**

Figure 1 shows the structures of the seven tin sulfide compounds examined in this study after structural optimisation with PBEsol.

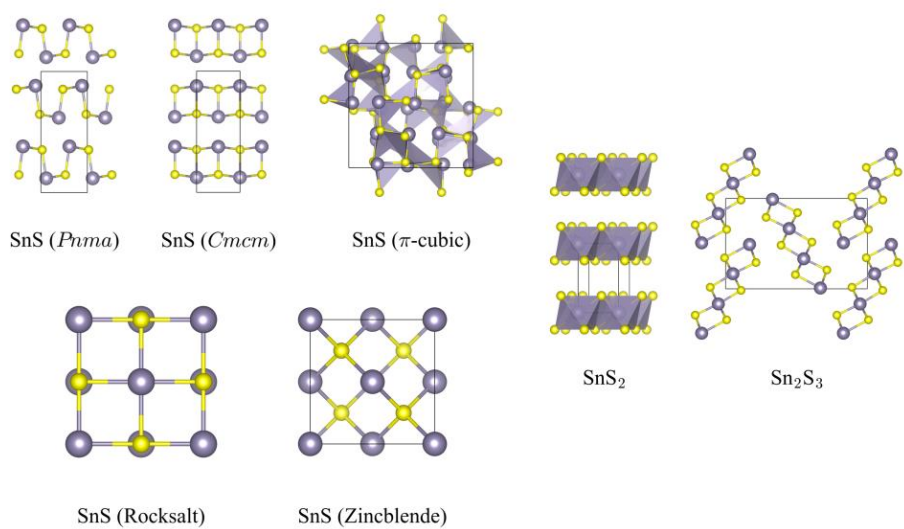

**Figure 1** PBEsol-optimised structures of the *Pnma*, *Cmcm*, π-cubic, rocksalt and zincblende phases of SnS, $SnS_2$ and $Sn_2S_3$. These images were prepared using the VESTA software.[44]

The structural diversity in the $Sn_xS_y$ phase space is primarily driven by the ability of Sn to adopt two different oxidation states, *viz.* Sn(II) and Sn(IV).

The *Pnma*, *Cmcm* and π-cubic phases of SnS are built up from distorted Sn(II) tetrahedra with a stereochemically-active lone pair occupying one of the four coordination sites. The two orthorhombic phases are composed of layered 2D SnS sheets, with the lone pairs projecting into the interlayer space and facilitating a weak interaction along the crystallographic *b* axes. The low-temperature *Pnma* phase can be thought of as a small distortion of the more symmetric high-temperature *Cmcm* phase, and a phase transition between them occurs above 875 K.[45] The local coordination in the π-cubic phase is similar to that in the orthorhombic phases, but the Sn-S bonding in this structure instead forms a 3D network. In the higher-symmetry rocksalt and zincblende structures, the Sn lone pair must be accommodated within the close-packed structure, and the stability of this arrangement depends sensitively on the local chemical environment of the cation.[46]

In $SnS_2$, symmetric edge-sharing Sn(IV) octahedra form 2D layers separated by a van der Waals' gap. $Sn_2S_3$ is a mixed oxidation-state phase containing equal proportions of Sn(II) and Sn(IV), and as such displays structural motifs from both SnS and $SnS_2$. The $Sn_2S_3$ structure consists of 1D chains of Sn(IV) octahedra capped by tetrahedral Sn(II), with weak interchain interactions facilitated by the Sn(II) lone pairs.

The optimised lattice parameters obtained with both exchange-correlation functionals (Table 2) were found to agree well with experimental measurements, and the PBEsol results were a good match to previous all-electron DFT calculations using the same functional.[5] With the exception of zincblende SnS, the dispersion correction consistently yielded a smaller lattice volume, with a particularly notable contraction in the non-bonded *c* direction of $SnS_2$. The reduction in volume implies that the attractive part of the dispersion correction is dominant for most of the structures, whereas for the zincblende phase the dispersion correction produces a net repulsive effect. This can be explained by the interaction of the Sn(II) lone pair with the symmetric coordination environment: the Sn-S bond length in the rocksalt structure (PBEsol: 2.874 Å, PBEsol + D3: 2.856 Å) is considerably longer than in the zincblende structure (PBEsol: 2.778 Å, PBEsol + D3: 2.780 Å), which would lead to a stronger (repulsive) interaction between the lone pair and the bonding electrons.

The formation energies obtained with both functionals, referenced to solid S and *β*-Sn, are compared in Table 3. These values were used to draw a convex hull including the newly-discovered π-cubic phase (Figure 2).

|  | PBEsol | | | PBEsol + D3 | | | Expt. | | |
| --- | --- | --- | --- | --- | --- | --- | --- | --- | --- |
|  | $a$ [Å] | $b$ [Å] | $c$ [Å] | $a$ [Å] | $b$ [Å] | $c$ [Å] | $a$ [Å] | $b$ [Å] | $c$ [Å] |
| α-Sn | 6.538 | - | - | 6.506 | - | - | 6.491[35] | - | - |
| β-Sn | 5.829 | - | 3.163 | 5.810 | - | 3.143 | 5.820[35] | - | 3.175[35] |
| S | 10.556 | 12.942 | 24.408 | 10.075 | 12.493 | 23.966 | 10.437[34] | 12.845[34] | 24.369[34] |
| SnS (*Pnma*) | 4.250 | 11.082 | 3.978 | 4.220 | 10.976 | 3.958 | 4.33[8] | 11.18[8] | 3.98[8] |
| SnS (Cubic) | 11.506 | - | - | 11.405 | - | - | 11.603[15] | - | - |
| SnS (*Cmcm*)[a] | 4.037 | 11.282 | 4.039 | 4.018 | 11.186 | 4.009 | 4.148[9] | 11.480[9] | 4.177[9] |
| SnS (Rocksalt) | 5.747 | - | - | 5.712 | - | - | - | - | - |
| SnS (Zincblende) | 6.416 | - | - | 6.420 | - | - | - | - | - |
| SnS$_2$ | 3.651 | - | 6.015 | 3.639 | - | 5.721 | 3.638[36] | - | 5.880[36] |
| Sn$_2$S$_3$ | 8.811 | 3.766 | 13.813 | 8.633 | 3.760 | 13.663 | 8.878[37] | 3.751[37] | 14.020[37] |

**Table 2** Optimised lattice constants of the models considered in this work obtained with the PBEsol and PBEsol + D3 exchange-correlation functionals. Experimental values are given where available for comparison. [a]The lattice constants in Ref. 9 are reported at 905 K, and are therefore expected to include significant thermal expansion.

Both functionals predict the groundstate of SnS to be the *Pnma* phase, with the high-temperature *Cmcm* phase calculated to be 1.76 kJ mol$^{-1}$ per SnS formula unit (F.U.; PBEsol) and 1.63 kJ mol$^{-1}$ per F.U. (PBEsol + D3) higher in energy. The π-cubic phase is metastable with respect to this equilibrium, being 2.19 kJ mol$^{-1}$ per F.U. higher in energy with PBEsol, and 2.54 kJ mol$^{-1}$ per F.U. higher in energy with PBEsol + D3. The rocksalt phase is around 4× higher in energy than the π-cubic phase, at 8.54 kJ mol$^{-1}$ per F.U. with PBEsol and 10.75 kJ mol$^{-1}$ per F.U. with PBEsol + D3. The zincblende phase has a much smaller formation energy, and is thus very high in energy relative to all the other phases (PBEsol: 72.23 kJ mol$^{-1}$ per F.U., PBEsol + D3: 80.55 kJ mol$^{-1}$ per F.U.).

As in other calculations,[6, 20] PBEsol predicts Sn$_2$S$_3$ to be above the convex hull, making it unstable with respect to decomposition into *Pnma* SnS and SnS$_2$.

| System | $E_F$ [kJ mol$^{-1}$ per F.U.] | | | Bulk Dynamically Stable? |
| --- | --- | --- | --- | --- |
| | PBEsol | PBEsol + D3 | Expt. | |
| SnS (*Pnma*) | -90.59 | -95.00 | -100 to -108 [47, 48] | Yes |
| SnS (π-cubic) | -88.40 | -92.46 | - | Yes |
| SnS (*Cmcm*) | -88.83 | -93.36 | - | No |
| SnS (Rocksalt) | -82.05 | -84.24 | - | No |
| SnS (Zincblende) | -18.36 | -14.45 | - | No |
| SnS$_2$ | -120.00 | -127.47 | -148 to -182 [48-50] | Yes |
| Sn$_2$S$_3$ | -208.90 | -222.32 | -249 to -297 [48-50] | Yes |

**Table 3** Energetic and dynamical stability of the *Pnma*, π-cubic, *Cmcm*, rocksalt and zincblende phases of SnS, SnS$_2$ and Sn$_2$S$_3$. The calculated formation energies ($E_F$) of each compound are compared to experimental data where available, while the last column lists the bulk dynamical stabilities assessed from the harmonic phonon dispersions (see Figure 3).

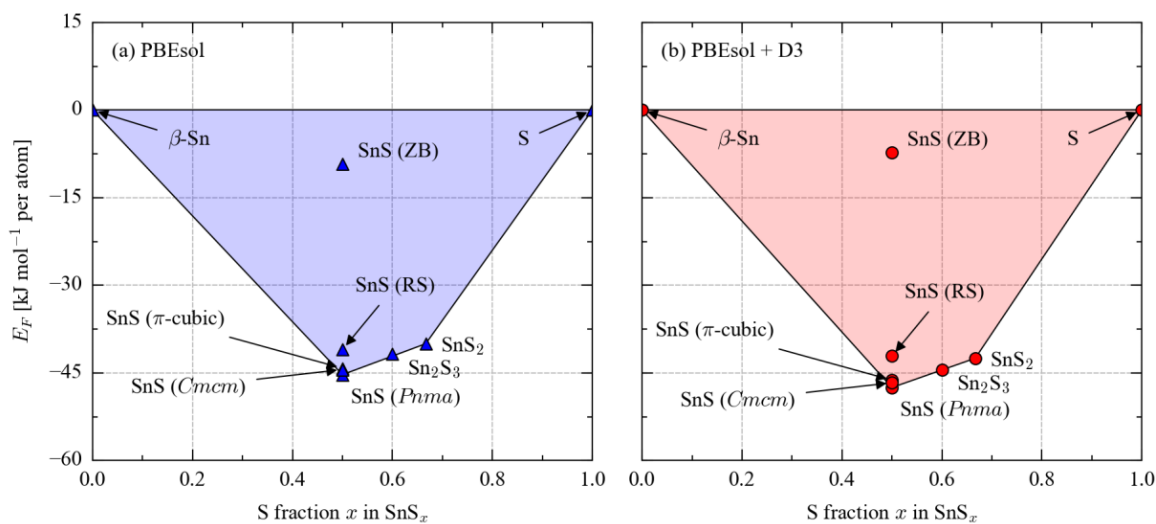

**Figure 2** Thermodynamic convex hulls showing the calculated phase stability of the seven compounds in the Sn$_x$S$_y$ system, based on formation energies obtained with (a) PBEsol and (b) PBEsol + D3. ZB and RS refer to the zincblende and rocksalt phases of SnS. The optimised crystal-structure parameters are reported in Table 2.

Whereas PBEsol predicts it to be 1.68 kJ mol$^{-1}$ above the hull, PBEsol + D3 predicts it to be much closer at 0.14 kJ mol$^{-1}$; this suggests that the dispersion correction selectively stabilises the sesquisulfide, most likely due to its improved description of the weak interactions between the bonded chains. Both values, but in particular the PBEsol + D3 one, are on the order of differences in the vibrational zero-point energy, a point which we return to below.

Finally, we note that the similar calculations carried out previously,[5] although not including the π-cubic phase, were referenced to *α*-Sn; in the present calculations, both functionals predicted *β*-Sn to be lower in energy, but a similar comparison of the formation energies and a convex hull referenced to *α*-Sn yielded the same qualitative results (Table S1 and Figure S1 in the supporting information).

We next consider the dynamical stability of the seven sulfides. In the harmonic phonon model, displacement of atoms from their equilibrium positions along a normal-mode coordinate $Q$ leads to a change in potential energy $U = \frac{1}{2}Q^2\omega^2$, where $\omega$ is the vibrational frequency. If a structure is a minimum on the potential-energy surface, $U$ and $\omega$ are $\geq 0$ (the three acoustic modes have $\omega = 0$ at the centre of the Brillouin zone (Γ)). If, on the other hand, a structure is a potential-energy maximum (e.g. a saddle point), atomic motion along one or more vibrations will lead to a lowering of the energy, and hence to a complex (imaginary) frequency.

In the present calculations, an analysis of the phonon dispersions (Table 1, Figure 3), that is, the frequencies of the 3*N* phonon bands as a function of the phonon wavevector, indicates the *Pnma* and π-cubic phases of SnS to be dynamically stable, whereas the *Cmcm*, rocksalt and zincblende phases all have imaginary modes.

The instability of the high-temperature *Cmcm* phase is anticipated, as the isostructural *Cmcm* phase of SnSe exhibits a similar phonon dispersion with imaginary modes at the same reciprocal-space wavevectors.[51] The *Cmcm* structure is effectively a low-lying saddle point on the potential-energy surface, connecting equivalent distorted *Pnma* minima. Above the phase-transition temperature, sufficient thermal energy is available for the system to rapidly "hop" between the minima, and the higher-symmetry phase is observed as a crystallographic average structure with significant local distortions. The implication of this is that the *Cmcm* phase is only observed above the phase-transition temperature, and cannot be isolated under ambient conditions, e.g. by rapid quenching.

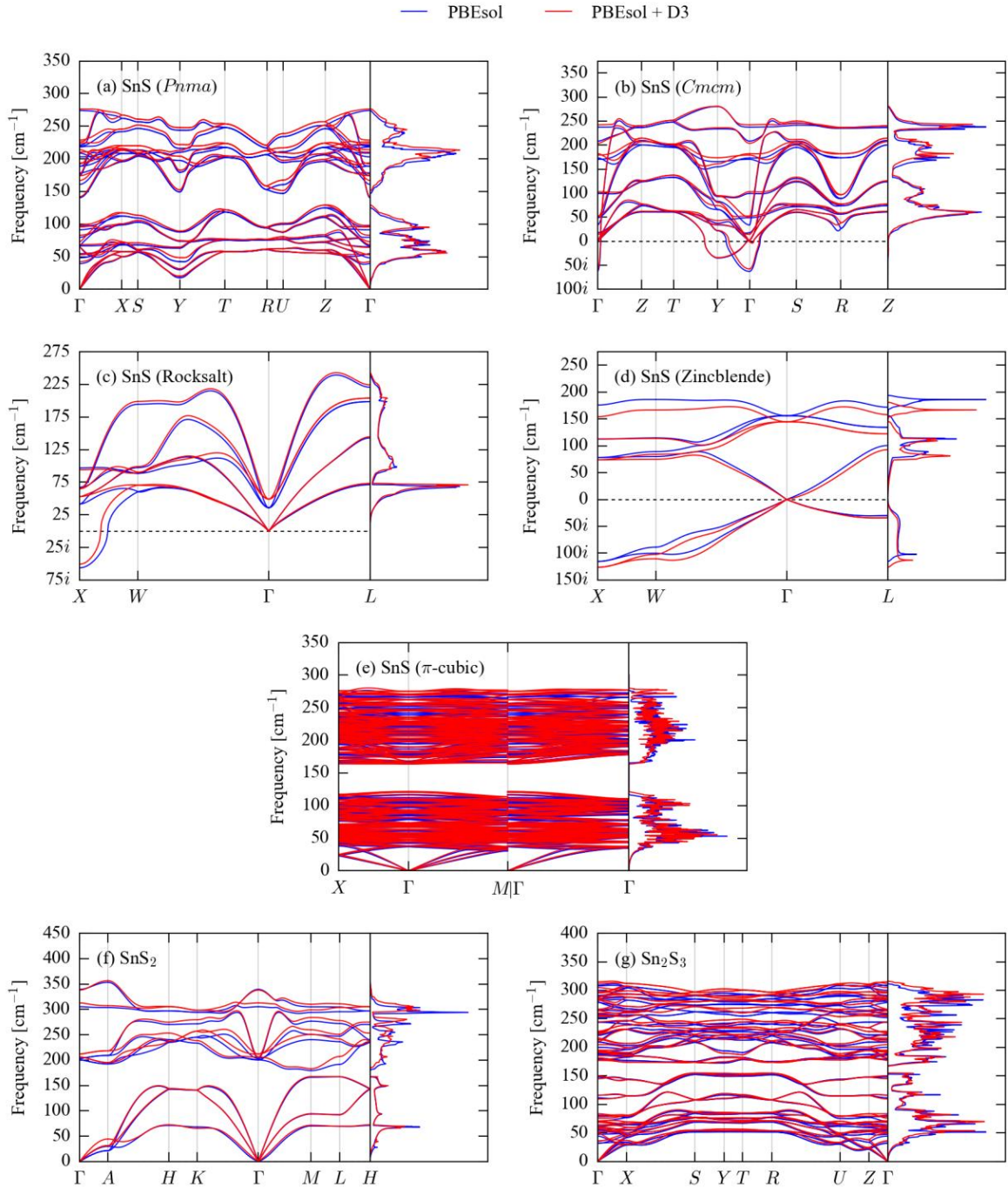

**Figure 3** Phonon dispersion and density of states curves for the *Pnma* (a), *Cmcm* (b), rocksalt (c), zincblende (d) and π-cubic (e) phases of SnS, $SnS_2$ (f) and $Sn_2S_3$ (g), calculated within density-functional theory with electron exchange and correlation described by PBEsol (blue) and PBEsol + D3 (red).

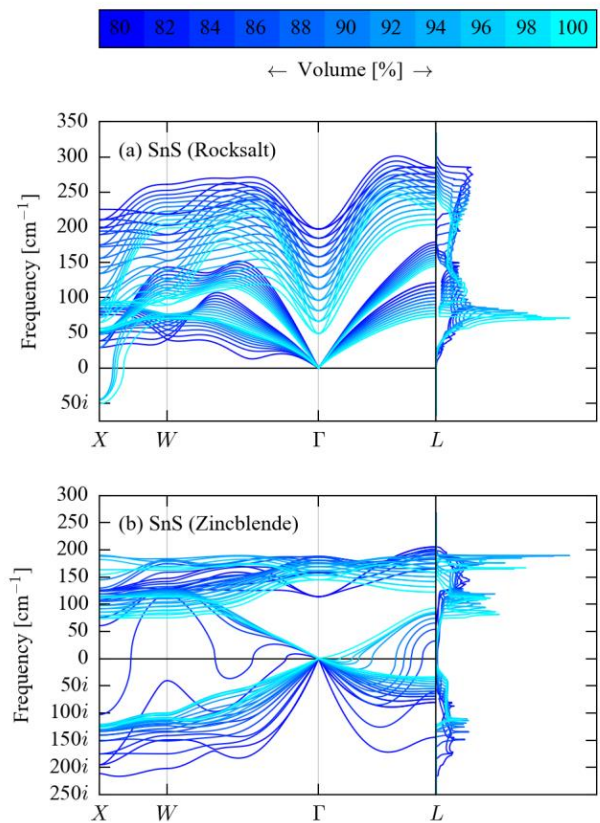

**Figure 4** Phonon dispersion and density of states curves for the rocksalt (a) and zincblende (b) phases of SnS calculated at cell volumes ranging from the athermal equilibrium (cyan) to a 20 % compression (blue). Both sets of curves were computed with PBEsol + D3, and the results from identical calculations with PBEsol are presented in Figure S2 (supporting information).

The phonon instabilities in the rocksalt phase correspond to a fourfold enlargement of the two-atom primitive cell, suggesting a transition towards the *Pnma* phase. Whereas the imaginary modes in this structure (and also in the *Cmcm* phase) are restricted to well-defined phonon wavevectors and represent a small fraction of the overall phonon density of states (DoS), the zincblende instabilities account for a significant proportion (~⅓) of the integrated phonon DoS. Taken together with its predicted high formation energy, this phase is extremely unlikely to form, even under non-ambient conditions such as high pressure.

To confirm this, we calculated phonon dispersions of the rocksalt and zincblende phases for volume compressions of up to 20 %, corresponding to a 0.41-0.46 Å reduction in the lattice constants (Figure 4, see also Figure S2 in the supporting information). Under moderate compression, the imaginary modes in the rocksalt dispersion harden, indicating that the bulk phase can be dynamically stabilised under pressure. In contrast, compression of the zincblende phase was found to induce further phonon softening.

These results strongly suggest that cubic SnS has been misassigned as zincblende in various past studies. For example, the cubic lattice constant reported in Ref. [13], 5.846 Å, is much closer to those of our rocksalt models (PBEsol: 5.747 Å, PBEsol + D3: 5.712 Å) than to those of our zincblende ones (PBEsol: 6.416 Å, PBEsol + D3: 6.420 Å). The rocksalt lattice constants are also a better match for the lattice constant of 6.00 Å measured for SnS films grown heteroepitaxially on NaCl. While it is possible that a rocksalt phase could be grown on certain substrates (or under pressure), we would argue that bulk cubic SnS is more likely to be the π-cubic phase, as: (1) our calculations show that this is dynamically stable in bulk; (2) the tetrahedral coordination environment in this structure is more likely to produce the tetrahedral morphology commonly observed in SnS nanoparticles;[11, 16] and (3) it has been shown that the powder X-ray diffraction pattern of this phase may have been misinterpreted as that of a zincblende phase in the past.[15, 17]

Both $SnS_2$ and $Sn_3S_3$ were found to be dynamically stable, as is expected given that these phases can be prepared as bulk single crystals.[18]

Comparing the phonon dispersion and DoS curves calculated using PBEsol and PBEsol + D3 reveals a general shift of the latter to higher frequencies due to the smaller predicted unit-cell volumes.[32] The exception is zincblende SnS, where the slightly larger lattice constant predicted with PBEsol + D3 leads to a corresponding softening of the phonon frequencies. For all seven compounds, however, the main features, in particular the presence or absence of imaginary modes, are qualitatively similar.

As well as assessing dynamical stability, lattice-dynamics calculations also allow vibrational contributions to the free energy to be taken into account when assessing thermodynamic stability. Within the harmonic model, the constant-volume (Helmholtz) free energy, $A$, is given by:

$$A(T) = U_{\text{latt}} + U_{\text{vib}}(T) - TS_{\text{vib}}(T) \tag{1}$$

where $U_{\text{latt}}$ is the (athermal) lattice energy (i.e. the total energy calculated using DFT, in the present study), $U_{\text{vib}}$ is the vibrational internal energy, and $S_{\text{vib}}$ is the vibrational entropy. $A(T)$ is typically evaluated from the lattice energy and phonon frequencies using the bridge relation from statistical mechanics:

$$\begin{aligned} A(T) &= -k_B T \ln Z(T) \\ &= U_{\text{latt}} + \frac{1}{N}\left\{\frac{1}{2}\sum_{\mathbf{q}v}\hbar\omega(\mathbf{q}v) + k_B T \sum_{\mathbf{q}v}\ln\left[1 - e^{\frac{-\hbar\omega(\mathbf{q}v)}{k_B T}}\right]\right\} \end{aligned} \tag{2}$$

where $Z(T)$ is the thermodynamic partition function, $k_B$ is Boltzmann's constant, the phonon frequencies $\omega$ are indexed by a reciprocal-space wavevector $\mathbf{q}$ and a band index $v$, and the sum over the phonon Brillouin zone is normalised by the number of unit cells in the crystal, $N$, equivalent to the number of wavevectors included in the summation.

The Helmholtz energies of the *Cmcm*, π-cubic and rocksalt phases of SnS, relative to the *Pnma* phase, are shown as a function of temperature in Figure 5, and free-energy differences calculated at 0, 300 and 900 K are listed in Table 4.

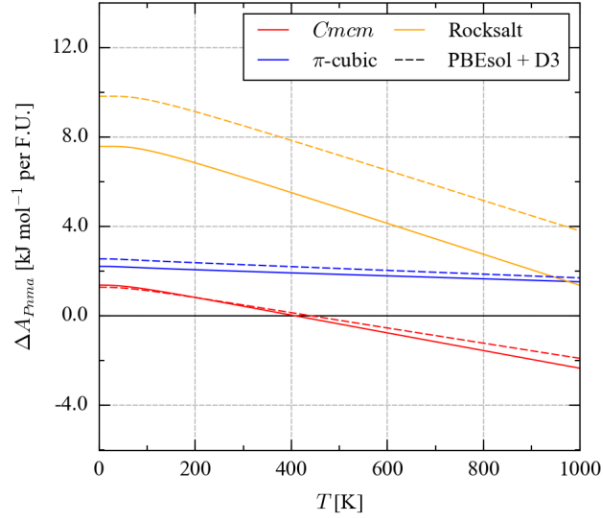

**Figure 5** Thermodynamic stability of the *Cmcm* (red), π-cubic (blue) and rocksalt (yellow) phases of SnS, as a function of temperature, relative to the low-temperature groundstate *Pnma* phase, based on the constant-volume (Helmholtz) free energies, *A*, calculated with PBEsol (solid lines) and PBEsol + D3 (dashed lines).

|  | [kJ mol$^{-1}$ per F.U.] | | | | | | | |
|---|---|---|---|---|---|---|---|---|
|  | PBEsol | | | | PBEsol + D3 | | | |
|  | $\Delta U_{latt}$ | $\Delta A_{0K}$ | $\Delta A_{300K}$ | $\Delta A_{900K}$ | $\Delta U_{latt}$ | $\Delta A_{0K}$ | $\Delta A_{300K}$ | $\Delta A_{900K}$ |
| *Pnma* | 0.00 | 0.00 | 0.00 | 0.00 | 0.00 | 0.00 | 0.00 | 0.00 |
| Cubic | 2.19 | 2.23 | 2.01 | 1.61 | 2.54 | 2.56 | 2.30 | 1.79 |
| *Cmcm* | 1.76 | 1.39 | 0.45 | -1.93 | 1.63 | 1.29 | 0.49 | -1.55 |
| Rocksalt | 8.54 | 7.63 | 6.24 | 2.11 | 10.75 | 9.85 | 8.53 | 4.51 |
| Zincblende | 72.23 | - | - | - | 80.55 | - | - | - |

**Table 4** Relative energies of the five phases of SnS compared to the groundstate *Pnma* phase. The lattice energies, $\Delta U_{latt}$, and constant-volume (Helmholtz) free energies, $\Delta A$, at 0, 300 and 900 K calculated with PBEsol and PBEsol + D3 are compared. Energies are given in kJ mol$^{-1}$ per SnS formula unit. Due to the large proportion of imaginary phonon modes in its phonon density of states, we did not calculate free energies for the zincblende phase.

Although, as noted above, the high-temperature *Cmcm* phase is an average structure, since the imaginary modes form a small part of the overall phonon DoS we make the approximation that the static configuration provides a reasonable representation of its (average) internal energy and lattice dynamics. Doing so predicts the free energies of the *Pnma* and *Cmcm* phases to cross at around 400-450 K, which, although qualitatively correct, is considerably lower than the 875 K at which the phase transition is observed experimentally. This can be ascribed to the approximations made in the present calculations, including the approximate treatment of the physics of the phase transition, and possibly also the neglect of volume expansion and other forms of anharmonicity.[51]

The π-cubic phase is predicted to remain metastable up to 1000 K, with an approximately constant free energy difference with respect to the *Pnma* phase. In contrast, the free energy of the rocksalt phase, computed under the same assumptions as for the *Cmcm* phase, falls with temperature, but remains above the *Pnma* phase in energy across the range of temperatures considered.

Finally, we also consider the effect of vibrational contributions to the free energy on the stability of $Sn_2S_3$. Figure 6 compares the temperature dependence of the free energy, $\Delta A$, of the disproportionation reaction:

$$Sn_2S_3 \text{ (s)} \rightarrow SnS_2 \text{ (s)} + SnS \text{ (s)} \quad\quad\quad (3)$$

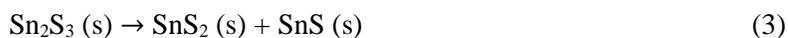

In light of the free-energy crossing evident in Figure 5, we consider decomposition to both *Pnma* and *Cmcm* SnS. The calculated reaction energies at representative temperatures are given in Table 5.

PBEsol and PBEsol + D3 both predict the decomposition to the *Pnma* phase to become less favourable with temperature (i.e. the free energy increases), whereas decomposition to the *Cmcm* phase becomes more favourable (i.e. the energy decreases). The PBEsol free energies predict decomposition to the *Pnma* phase to become unfavourable above ~650 K; however, decomposition to the *Cmcm* phase is a lower-energy pathway above ~400 K, and the reaction remains favourable.

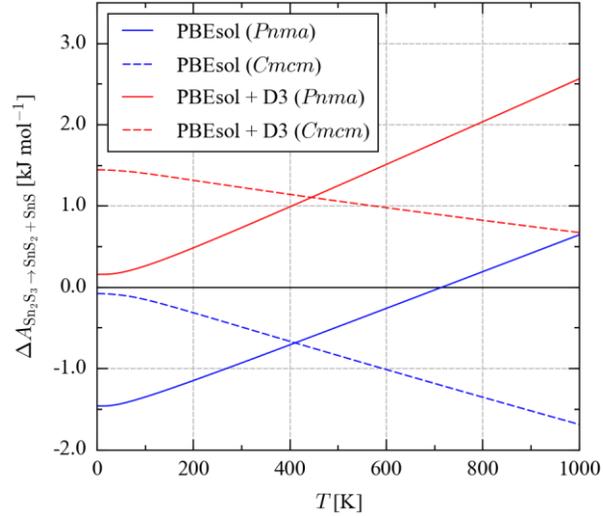

**Figure 6** Constant-volume (Helmholtz) free energy, $A$, for the decomposition of $Sn_2S_3$ into $SnS_2$ and SnS as a function of temperature, calculated with PBEsol (blue) and PBEsol + D3 (red). Based on the stability ordering of the SnS phases (Figures 2/5, Tables 3/4), we consider decomposition to both the *Pnma* (solid lines) and *Cmcm* (dashed lines) phases of SnS.

|  | [kJ mol$^{-1}$ Per F.U.] | |
| --- | --- | --- |
|  | PBEsol | PBEsol + D3 |
| $\Delta U_{latt}$ | -1.68 | -0.14 |
| $\Delta A_{0K}$ | -1.51 | 0.12 |
| $\Delta A_{300K}$ | -0.98 | 0.69 |
| $\Delta A_{900K}$ | 0.37 | 2.26 |
| *Cmcm* SnS: $\Delta A_{900K}$ | -1.57 | 0.71 |

**Table 5** Reaction energies for the decomposition reaction $Sn_2S_3 \rightarrow SnS_2 + SnS$, calculated based on the differences in the athermal lattice energies ($\Delta U_{latt}$) and the constant-volume (Helmholtz) free energies ($\Delta A$) at 0, 300 and 900 K. A negative energy implies that $Sn_2S_3$ is unstable, while a positive energy indicates the reverse. The energies are given with respect to *Pnma* SnS, as this was determined to be the energetic ground-state, while the 900 K free energy calculated with respect to the *Cmcm* phase, which has the lowest free energy of the SnS phases above ~400 K, is also given.

With PBEsol + D3, on the other hand, the inclusion of zero-point energy renders decomposition unfavourable at 0 K, and it remains so up to 1000 K.

Given the theoretical reports predicting $Sn_2S_3$ to be unstable with respect to *Pnma* SnS and $SnS_2$, it is of interest to ascertain whether this result can also be obtained with other dispersion correction methods. We therefore performed calculations on *Pnma* and *Cmcm* SnS, $SnS_2$ and $Sn_2S_3$ with the newer variant of PBEsol + D3 with the Beck-Johnson damping scheme[52] (see Table S2 and Figures S3/S4 in the supporting information). These calculations yielded the same qualitative result, i.e. that the dispersion correction brings $Sn_2S_3$ closer to the convex hull than uncorrected PBEsol, and the subsequent inclusion of the vibrational contributions to the free energy predicts it to be thermodynamically stable up to 1000 K. Although the Becke-Johnson-damped DFT-D3 is similar to the "bare" D3, we note that the majority of the other well-known dispersion corrections, e.g. DFT-D2,[53] the Tkatchenko-Scheffler method (DFT-TS),[54] and the more sophisticated many-body dispersion technique,[55, 56] do not have scaling constants optimised for PBEsol. To investigate other common dispersion corrections would thus also require the exploration of other GGA functionals, which we defer to a future study.

Although, as noted above regarding the predicted *Pnma*-to-*Cmcm* phase-transition temperature, these calculations may still be missing some important effects, the predicted stability of $Sn_2S_3$ with PBEsol + D3 is in keeping with experimental findings. These results therefore highlight the potential importance of both a sufficient description of dispersion interactions and of vibrational contributions to the free energy in assessing the phase stability of the tin sulfides.

**Conclusions**

In summary, this work has provided detailed insight into the thermodynamic and dynamical stability of the seven currently known and proposed compounds in the $Sn_xS_y$ phase space, addressing several key outstanding questions.

The recently-discovered π-cubic phase is metastable with respect to the orthorhombic *Pnma*/*Cmcm* equilibrium. The bulk rocksalt phase is higher in energy and is dynamically unstable, but could potentially be stabilised under pressure or epitaxial strain. Our calculations show conclusively that the hypothetical zincblende phase is both energetically and dynamically unstable, and we suggest that reports of this phase be reassessed as either of the other two cubic phases.

Finally, we also show that accurately modelling the tin sulfide phase diagram, in particular reproducing the experimentally-observed stability of the sesquisulfide, may require theoretical techniques which afford a good treatment of dispersion interactions, and possibly also the inclusion of contributions to the energetics from lattice dynamics. This result provides a baseline for further theoretical characterisation of this less-well-studied tin sulfide phase.

**Electronic supporting information**

Electronic supporting information includes formation energies and a convex hull calculated with reference to *α*-SnS, volume-dependent phonon dispersions for rocksalt and zincblende SnS calculated with the PBEsol functional, optimised lattice parameters and phonon dispersions of *Pnma* and *Cmcm* SnS, $SnS_2$ and $Sn_2S_3$ calculated with PBEsol + D3 (BJ), and the calculated temperature-dependent Helmholtz free energy of decomposition of $Sn_2S_3$ obtained with PBEsol + D3 (BJ).

**Data-access statement**

The data from these calculations, including the optimised structures, force constants, phonon dispersions and density of states curves, and thermodynamic functions, are available in an online repository at https://doi.org/10.15125/BATH-00339 (see also https://github.com/WMD-group/Phonons).

**Acknowledgements**

JMS gratefully acknowledges support from an EPSRC Programme Grant (grant no. EP/K004956/1). LAB is an International Research Fellow of the Japan Society of Promotion of Science (JSPS; grant no. 26.04792). AW acknowledges support from the Royal Society and the ERC (grant no. 277757). Calculations were carried out using the SiSu supercomputer at the IT Center for Science (CSC), Finland, via the Partnership for Advanced Computing in Europe (PRACE) project no. 13DECI0317/IsoSwitch, and on the Balena HPC cluster at the University of Bath, which is maintained by Bath University Computing Services. Some of the calculations were also carried out on the UK national Archer HPC facility, accessed through membership of the UK Materials Chemistry Consortium, which is funded by EPSRC grant no. EP/L000202.

**TOC Graphic**

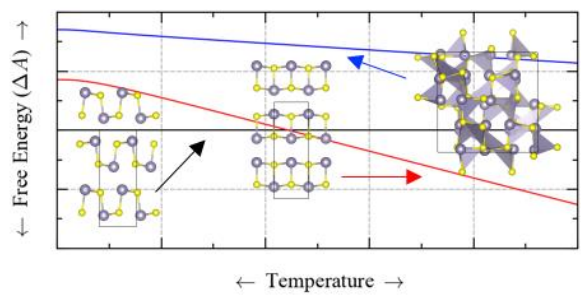

# Supporting Information:

# Chemical and Lattice Stability of the Tin Sulfides

Jonathan M. Skelton,[1,*] Lee A. Burton,[2] Fumiyasu Oba,[2] and Aron Walsh[1,3,4]

[1] *Department of Chemistry, University of Bath, Claverton Down, Bath BA2 7AY, UK*

[2] *Laboratory for Materials and Structures, Institute of Innovative Research, Tokyo Institute of Technology, 4259 Nagatsuta, Midori-ku, Yokohama 226-8503, Japan*

[3] *Department of Materials, Imperial College London, Exhibition Road, London SW7 2AZ, UK*

[4] *Global E³ Institute and Department of Materials Science and Engineering, Yonsei University, Seoul 120-749, Korea*

[*] *Corresponding author; e-mail: j.m.skelton@bath.ac.uk*

| | $E_F$ [kJ mol⁻¹ per F.U.] | | |
|---|---|---|---|
| System | PBEsol | PBEsol + D3 | Expt. |
| SnS (*Pnma*) | -93.04 | -98.93 | -100 to -108[1,2] |
| SnS (π-cubic) | -90.85 | -96.39 | - |
| SnS (*Cmcm*) | -91.29 | -97.29 | - |
| SnS (Rocksalt) | -84.50 | -88.17 | - |
| SnS (Zincblende) | -20.81 | -18.37 | - |
| SnS$_2$ | -122.45 | -131.40 | -148 to -182[2-4] |
| Sn$_2$S$_3$ | -213.81 | -230.18 | -249 to -297[2-4] |

**Table S1** Formation energies ($E_F$) of the *Pnma*, π-cubic, *Cmcm*, rocksalt and zincblende phases of SnS, SnS$_2$ and Sn$_2$S$_3$, calculated with reference to *α*-Sn. These may be compared to the data in Table 3 in the text. Experimental values are given where available for comparison.

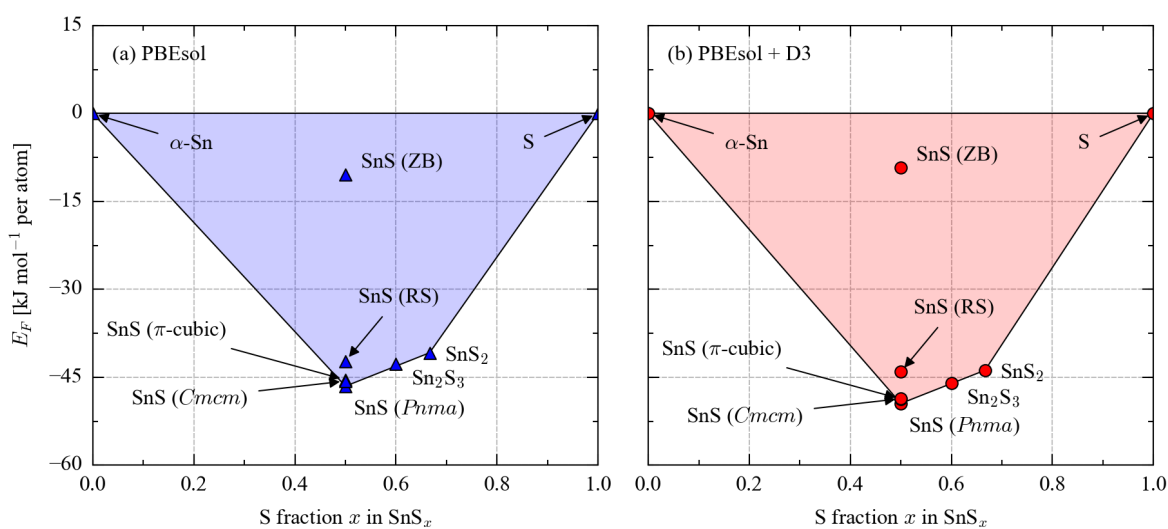

**Figure S1** Convex hulls calculated for the Sn$_x$S$_y$ system, based on the formation energies in Table S1 (i.e. referenced to *α*-Sn) obtained with (a) PBEsol and (b) PBEsol + D3. These may be compared to the convex hulls presented in Figure 2 in the text.

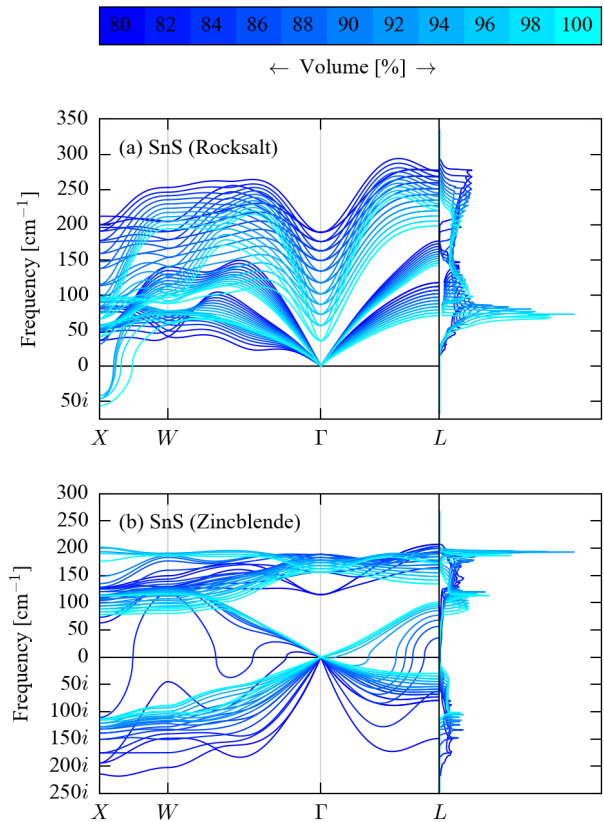

**Figure S2** Phonon dispersion and density of states curves for the rocksalt (a) and zincblende (b) phases of SnS calculated at cell volumes ranging from the athermal equilibrium (cyan) to a 20 % compression (blue). Both sets of curves were computed with PBEsol, and may be compared to the corresponding plots in Figure 4 in the text.

- Page S3 -

|  | PBEsol | | | PBEsol + D3 | | | PBEsol + D3 (BJ) | | | Expt. | | |
| --- | --- | --- | --- | --- | --- | --- | --- | --- | --- | --- | --- | --- |
|  | $a$ [Å] | $b$ [Å] | $c$ [Å] | $a$ [Å] | $b$ [Å] | $c$ [Å] | $a$ [Å] | $b$ [Å] | $c$ [Å] | $a$ [Å] | $b$ [Å] | $c$ [Å] |
| SnS (*Pnma*) | 4.250 | 11.082 | 3.978 | 4.220 | 10.976 | 3.958 | 4.113 | 10.887 | 3.953 | 4.33[5] | 11.18[5] | 3.98[5] |
| SnS (*Cmcm*)[a] | 4.037 | 11.282 | 4.039 | 4.018 | 11.186 | 4.009 | 3.987 | 11.005 | 3.988 | 4.148[6] | 11.480[6] | 4.177[6] |
| $SnS_2$ | 3.651 | - | 6.015 | 3.639 | - | 5.721 | 3.625 | - | 5.584 | 3.638[7] | - | 5.880[7] |
| $Sn_2S_3$ | 8.811 | 3.766 | 13.813 | 8.633 | 3.760 | 13.663 | 8.420 | 3.754 | 13.532 | 8.878[8] | 3.751[8] | 14.020[8] |

**Table S2** Lattice parameters of *Pnma* and *Cmcm* SnS, $SnS_2$ and $Sn_2S_3$ obtained with the PBEsol, PBEsol + D3 and PBEsol + D3 (BJ) exchange-correlation functionals. The PBEsol and PBEsol + D3 values data are reproduced from Table 2 in the text for comparison. Experimental values are also given for comparison. [a]The lattice constants in Ref. 6 are reported at 905 K, and are therefore expected to include significant thermal expansion.

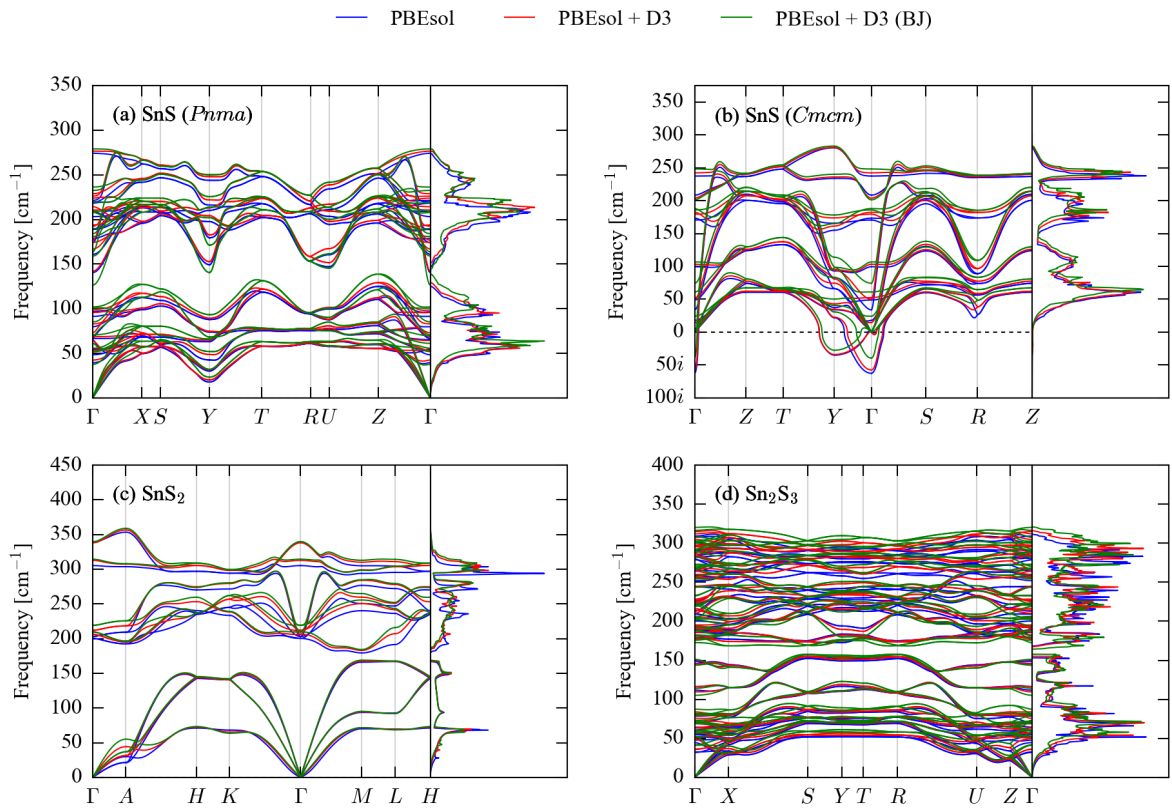

**Figure S3** Phonon dispersion and density of states curves for *Pnma* (a) and *Cmcm* (b) SnS, $SnS_2$ (c) and $Sn_2S_3$ (d), calculated using the PBEsol (blue), PBEsol + D3 (red) and PBEsol + D3 (BJ) (green) functionals. The PBEsol and PBEsol + D3 data are duplicated from the relevant subplots of Figure 3 in the text for comparison.

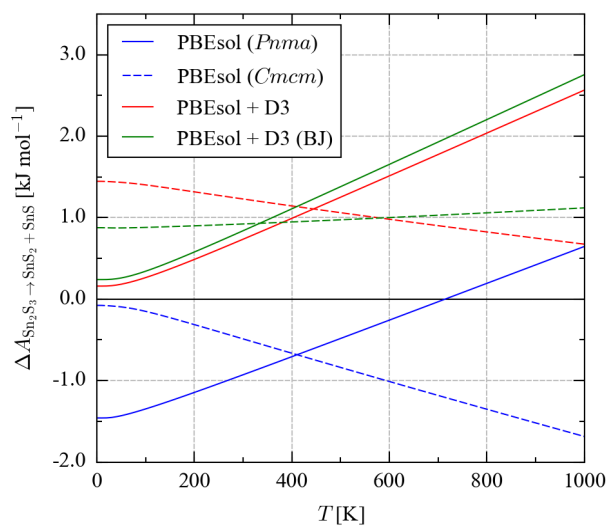

**Figure S4** Constant-volume (Helmholtz) free energy for the decomposition reaction $Sn_2S_3 \rightarrow SnS_2 + SnS$ as a function of temperature, calculated with PBEsol (blue), PBEsol + D3 (red) and PBEsol + D3 (BJ) (green). Based on the stability ordering of the SnS phases (Figures 2/5 and Tables 3/4 in the text), we consider decomposition to both the *Pnma* (solid lines) and *Cmcm* (dashed lines) phases of SnS.